\documentclass[twocolumn,showpacs,prb]{revtex4}

\usepackage{dcolumn}
\usepackage{bm}
\usepackage{amsmath,amsthm}
\usepackage{amssymb}
\usepackage{graphicx}
\usepackage{color}
\usepackage{enumerate}

\newlength{\minipagewidth}
\begin{document}

\preprint{APS/}

%
%
  
\title{Reducing the thermal conductivity of carbon nanotubes below the random isotope limit}

\author{Gabriel Stoltz}
\affiliation{
  Universit\'e Paris Est, CERMICS,
  Project-team MICMAC, INRIA-Ecole des Ponts, 
  6 \& 8 Av. Pascal, 77455 Marne-la-Vall\'ee Cedex 2, France
}

\author{Natalio Mingo}
\affiliation{
  LITEN, CEA Grenoble, 17 rue des Martyrs, 38000 Grenoble, France
}
\author{Francesco Mauri}
\affiliation{
  IMPMC, Universit\'es Paris 6 et 7, CNRS, IGPG, 140 rue de Lourmel,
  75015 Paris, France
}

\date{\today}

\begin{abstract}
We find that introducing segmented isotopic disorder patterns may
considerably reduce the thermal conductivity of pristine carbon nanotubes below
the uncorrelated disorder value. This is a result of the
interplay between different length scales in the phonon scattering
process. We use ab-initio atomistic Green's function calculations to
quantify the effect of various types of segmentation similar to that
experimentally produced by coalescence of isotope-engineered fullerenes.
\end{abstract}

\pacs{
65.80.+n, 
61.46.Fg, 
63.22.-m, 
63.20.kp 
}

\maketitle

%
%

\section{Introduction}

Recent experimental and theoretical works have shown that introducing
lattice matched nanocrystals into an alloy can further reduce its
thermal conductivity, in what has been termed "beating the alloy
limit"\cite{Kim,KimMajumdar,Mingo09}. This has a very important
practical application in the field of thermoelectric materials, in
which a major goal is to obtain low thermal conductivities without
affecting electronic properties. The fact that
clusters of scatterers can scatter phonons more efficiently than they
would if they acted separately has been known for some
time~\cite{Turk,Slack}. However, combining atomic and cluster
scattering effects in order to minimize the thermal conductivity had
not been clearly achieved until recently, and only for three
dimensional systems~\cite{Kim}.

In confined systems, such as nanowires or nanotubes, there appear to
be no experimental reports on this approach. The case of isotopic scattering in
carbon nanotubes is especially interesting in this respect. First, it
represents a well defined system, where theory~\cite{Savic,SLM09,Savic2} has been successful in
explaining measurements~\cite{chang06,chang08}. In addition, the use of isotopic
disorder means that electron transport will not be affected at
all. This means that the thermal conductivity could be tailored
independently of the electrical properties of the system, which is
especially attractive to nanoelectronics. In contrast, we do not think carbon nanotubes may be used for thermoelectric applications, due to their still too large thermal conductivity.

One problem is how to create nanostructures with isotopes. Isotopes of
different masses are chemically identical, so thermodynamically
speaking it is in principle not possible to arrange them inside the
solid into anything other than an uncorrelated alloy. Nevertheless,
there 
are ways to achieve
nanotubes where the isotopic concentration
is distributed in a segmented way. 
One of them consists in
synthesizing nanotubes from C$_{60}$ peapods~\cite{refC60bis,refC60}. In particular, 
C$_{60}$ buckyballs can be synthesized with different isotopic
concentrations. For example, pure $^{13}$C buckyballs, pure
$^{12}$C buckyballs, and $^{12}$C$_{0.5}$$^{13}$C$_{0.5}$ buckyballs could be
produced separately. These
different buckyball types can then be mixed together and get to form
nanopeapods, with the buckyballs encapsulated into nanotubes. Once
there, they can be heated up until they coalesce to form an inner
nanotube. In absence of carbon diffusion, 
the isotope concentration of the inner tube keeps a segmented
distribution in accordance with the distribution of the original
buckyball types. 
A detailed comparison between the measured and simulated Raman spectra
of heated nanopeapods 
has clearly demonstrated
that it is indeed
possible to obtain segmented isotope distributions~\cite{refC60}.
The outer nanotube that has been used to hold the
buckyballs can subsequently be removed by, e.g., burning the outer tube
with an electrical current~\cite{burning,burning2,burning3}.
A possible alternative route that may enable the growth of isotopically segmented carbon nanotubes is via chemical vapor deposition techniques \cite{Maruyama}.

In this work we study thermal transport in isotope engineered carbon nanotubes.
We show that the thermal conductivity can be significantly reduced below the 
random isotope alloy limit
in nanotubes produced from buckyball nanopeapods in which the isotopic disorder is 
nanostructured.

%
%
\section{Types of disorder}

\begin{figure}
\center
\includegraphics[width=7.0cm]{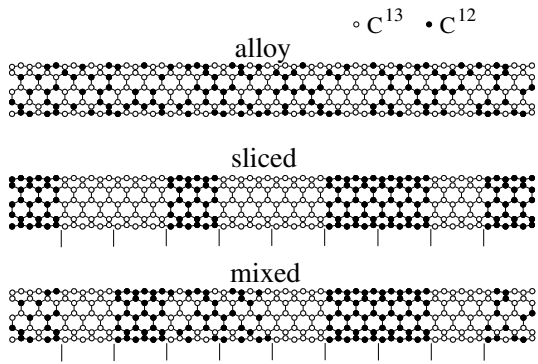}
\caption{\label{fig:types} 
Types of disorder. In the alloy case the isotopes are randomly distributed. In the
other two cases the isotope distribution is nanostructured.
The separation of homogeneous 60-atom sections is indicated by 
vertical ticks.
}
\end{figure}

We consider (5,5) armchair nanotubes, that have a diameter of 0.68 nm, 
close to that observed experimentally for the inner tubes produced by the coalescence of
buckyball peapods~\cite{refC60}. The unit cell of (5,5) tubes contains 20 atoms.
We consider three kinds of isotopic disorder, see Fig.~\ref{fig:types}.
In all the 3 cases we consider a 50\% mixture of $^{12}$C or of $^{13}$C atoms.
In the random {\it alloy} case the two isotopes are randomly distributed throughout atomic sites.
{\it Sliced} disorder corresponds to choosing at random 
whether a region of 60 consecutive atoms (3 tube cells)
will be composed of either $^{12}$C or of $^{13}$C atoms.
{\it Mixed} disorder
is a combination of both disorders, each 60-atom section being either 
homogeneously composed of $^{12}$C or $^{13}$C atoms, or 
of alloy type, mixing $^{12}$C and $^{13}$C in a random fashion.
In the latter case we assume that the probability to have an isotopically-pure or a
mixed section is the same. Finally, for sliced disorder, we also addressed the effect of 
section size (20, 60, 120 and 240 atoms) on the mean free paths.

%
%
\section{Computations}

To compute the phonon-transport properties, we 
use interatomic force constants 
obtained from density functional theory calculations,
as described in ~\cite{SLM09}.
The transmission functions $T_L(\omega)$ for a given length $L$ of the nanotube
and a phonon pulsation $\omega$ are computed using 
nonequilibrium Green's functions~\cite{SLM09}.
In this model, a disordered region is attached to two semi-infinite perfect leads,
which are kept at different temperatures, so that there is a net energy flow from
one end of the chain to the other.
The ballistic transmission $T_{\rm ballistic}(\omega)$, 
obtained in the case when there is no mass disorder, does not depend on the system length.

\begin{figure}
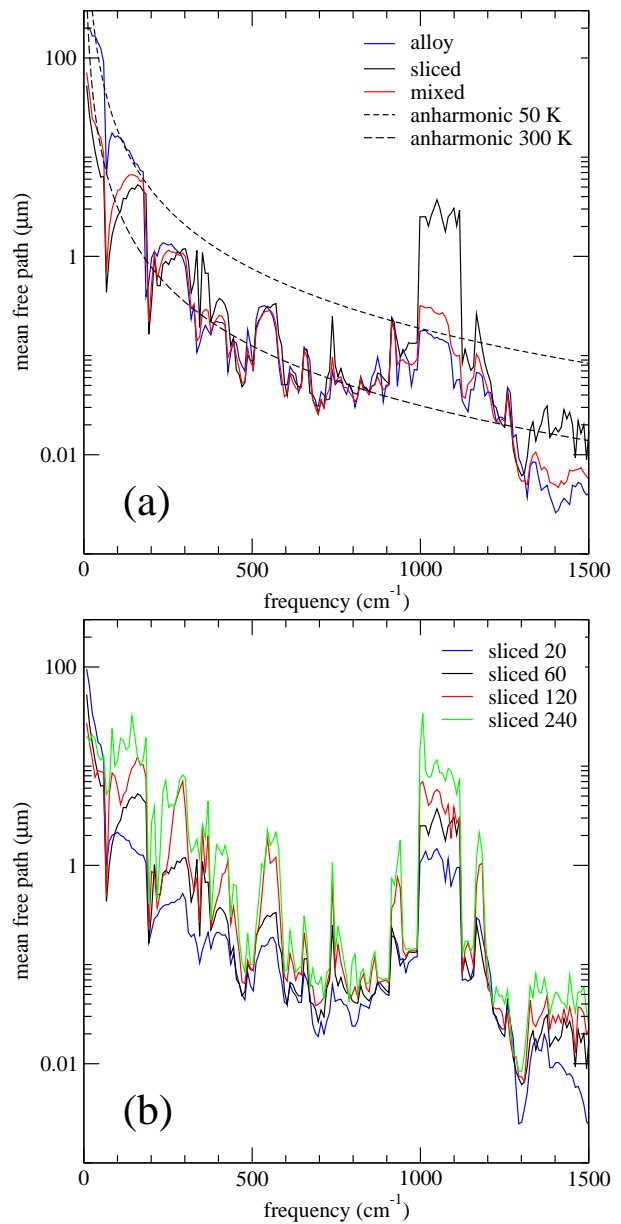

\center
\includegraphics[width=8.0cm]{fig2.eps}
\includegraphics[width=8.0cm]{fig2.5.eps}
\caption{\label{fig:harmonic_mfp} (color online)
 (a) Mean free paths $l_{\rm isotope}(\omega)$ for different types of isotopic disorder. The segments in the sliced case comprise 60 atoms.
 Dashed lines: anharmonic mean free path, $l_{\rm anh}(\omega)$, of pure nanotubes at 50 and 300K.
(b) Mean free path in the sliced case for different sizes (number of atoms) of the segments. See explanations in main text.
}
\end{figure}

The phonon mean free paths $l_{\rm isotope}(\omega)$ 
are obtained as described in Ref.~\cite{Savic2},
by computing short tube length transmissions, and assuming a diffusive behavior,
\textit{i.e.}
\begin{equation}
\label{eq:def_l_isotope}
T_L(\omega) = T_{\rm ballistic}(\omega) \frac{1}{1+L/l_{\rm isotope}(\omega)}.
\end{equation}
Therefore, 
\begin{equation}
\label{eq:lsq}
l_{\rm isotope}(\omega) = \left(\frac{T_{\rm ballistic}(\omega)}{T_L(\omega)}-1\right)^{-1} L.
\end{equation}
We estimated $l_{\rm isotope}$ by computing the transmission 
(averaged over 60 independent realizations of the 
mass disorder) as a function of the length, for short tubes of lengths 
$L = 7.5, 15, \dots, 75$~nm, and performing a least-square fit based on 
\eqref{eq:lsq}.
The estimated mean free paths are presented in Figure~\ref{fig:harmonic_mfp}. 
As a consistency check, we verified that the mean free paths obtained for the lowest frequency
modes are independent of the contact, by comparing results obtained for 
$^{13}$C atoms randomly inserted in an otherwise perfect $^{12}$C tube, and 
$^{12}$C atoms randomly inserted in an otherwise perfect $^{13}$C tube.
In general, random disorder
scatters short wavelengths (high frequencies) quite efficiently, but it does not affect
long wavelengths (
low frequencies) as strongly. On the other hand, larger
inhomogeneities comparable in size to the wavelength are much more
effective in scattering the low frequency phonons. This can be seen in
the plot of the mean free paths in Fig.~\ref{fig:harmonic_mfp}. At low frequency, sliced disorder yields
shorter mean free paths than the other disorder types. On the other hand, at high frequency it
has longer mean free paths than the other two types. In the sliced case, the figure also shows two clearly differentiated dependences with cluster size: at the lower frequencies, the larger the slice, the shorter the mean free path; for higher frequencies on the other hand, the shorter the slices, the shorter the mean free path. The latter behavior is a result of the larger density of interfaces when the slices are small. The low frequency phonons carry most of the
heat, due to their much longer mean free paths. Therefore, by employing sliced disorder, 
it is possible to decrease the 
thermal conductance of the nanotube below the random disorder value.

Figure~\ref{fig:harmonic_mfp} also shows an enhanced mean free path at a frequency window around 1100 cm$^{-1}$ for the segmented case. This is the result of angular momentum conservation which forbids many scattering processes. We verified that such resonance decreases dramatically if the symmetry of the segments is broken by making their edges irregular. This feature turns out to give a negligible contribution to the thermal conductance. We have checked that its presence or absence makes no difference in the final thermal conductance, which is determined by the much longer mean free paths of the lowest frequencies.

In order to estimate the temperatures at which the disorder pattern has an influence,
we combine, by Mathiessen's rule, the mean free paths computed above, and a rough estimate of
anharmonic relaxation length in graphite materials~\cite{KP94}:
$l_{\rm anh}(\omega) = B T^{-1} \omega^{-2}$,  
with 
$B = 9.38$~K$\cdot$ cm$^{-2}$m. We show this relaxation length as the dashed lines in Fig.~\ref{fig:harmonic_mfp}(a). 
The total relaxation length is therefore 
$l_{\rm diff}(\omega) = [l_{\rm anh}(\omega)^{-1} + l_{\rm isotope}(\omega)^{-1}]^{-1}$,
and an estimate of the transmission function for a CNT of length $L$ is obtained as
\begin{equation}
  \label{eq:transmission_anh}
  T_L(\omega) = T_{\rm ballistic}(\omega) \frac{1}{1+L/l_{\rm diff}(\omega)}.
\end{equation}
Once the transmission is known, the thermal conductance can be computed as \cite{MingoBroidoPRL05}
\[
g(L,T) = \int_{0}^{+\infty} \frac{\hbar \omega}{2\pi} T_L(\omega) 
  \, \frac{\partial f_{T}(\omega)}{\partial T} \, d\omega,
\]
where $f_T(\omega) = \left\{\exp\left[\hbar \omega/({k_\text{B}
  T})\right]-1 \right\}^{-1}$ is the Bose-Einstein distribution.

\begin{figure}
\center
\includegraphics[width=7.5cm]{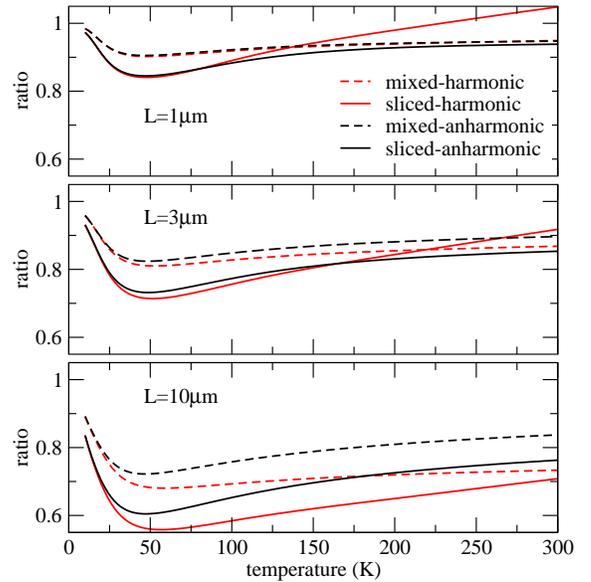}
\caption{\label{fig:anharmonic} (color online)
  Ratios of the thermal conductance of CNTs with mixed or sliced (60 atom segments) disorder, 
  divided by the thermal conductance of CNTs with alloy disorder, 
  as a function of the temperature, for different disorder patterns,
  and for tubes of lengths $L = $ 1, 3, 10~$\mu$m.
  Red lines represent results in the case when
  anharmonicity is not taken into account
  (using only Eq.~\eqref{eq:def_l_isotope}), and the black lines gives the 
  ratio when anharmonicity is present (using Eq.~\eqref{eq:transmission_anh}).
}
\end{figure}

Fig.~\ref{fig:anharmonic} compares the conductance computed
with mixed or sliced disorder, to the conductance obtained with random mass disorder,
as a function of temperature, for different lengths of the tube, with and without the
contribution of anharmonic scattering. The segments in the sliced case considered are 60 atoms in size.
Nanostructured disorder can reduce the thermal
conductivity of the nanotubes as much as 50\% below the isotope alloy
limit for low enough temperatures. 
Therefore, it is possible to obtain thermal
conductivities clearly lower than those of the random disorder
nanotube.

The observed reductions depend on the temperature and also on the
nanotube length. At temperatures high enough to have important
anharmonic scattering, isotope scattering becomes of secondary
importance, and nanostructured disorder has a smaller influence.
Similarly, if the nanotubes are very short, transport becomes
ballistic, and the effect of isotopes on the thermal conductance
becomes less apparent. Lowering the temperature increases the impact
of nanostructuring, down to a certain minimum at a particular
temperature. Below this temperature, the reduction decreases
because heat is increasingly carried by low frequency modes, which
have very long mean free paths. 
A general conclusion is that the effects should be
maximized for longer nanotubes at low temperatures.

%
%

\section{Discussion}

It should be possible to experimentally verify the predicted effects,
by synthesizing segmented disordered nanotubes in the way we proposed
earlier. Such nanostructured-disordered nanotubes could then be
suspended and contacted, and their thermal conductivity could be
measured using direct methods \cite{Shi,Pop}. 
Our calculations show that the effects should be
clearly noticeable in nanotubes 3 micrometers long at low temperature.
Thermal conductivity reductions of 50\% have been reported for BN nanotubes in
a wide range of temperatures. The phonon structures of BN and C nanotubes are rather similar, and 
the isotope effect at the same concentrations affects them in a similar way \cite{Savic}. Thus,
carbon nanotubes with 
sliced
disorder at 50\% isotope concentration can display even larger
reductions than those reported for BN. The additional effect of segmentation reported here implies that
it might be possible to reduce the low temperature (50K) thermal conductivity of long (10 $\mu$m) pristine carbon nanotubes to less than 
30\% of the isotopically pure value. Such strong effects are also suggested by the theoretical mean free paths in Fig.~\ref{fig:harmonic_mfp}.

%
%
\section{Conclusions}

In conclusion, nanostructured disorder leads to an important reduction
of thermal conductivity in carbon nanotubes. Reductions 40\% below the
random disorder case can be obtained for $^{12}$C$_{0.5}$$^{13}$C$_{0.5}$ 
nanotubes a few
micrometers long. Mixed disorder alternating alloy and pure sections
is less effective than pure sliced disorder. The relative reduction
increases with the nanotube length. For a given length, there
is a temperature that minimizes the thermal conductivity. For 
nanotubes a few micrometers long, this minimum occurs at cryogenic temperatures in the
20-100~K range. We have described a feasible way to produce
isotopic segmented nanotubes, which should allow for the
experimental verification of the results predicted here. The physical
phenomenon described here is not restricted to carbon nanotubes, and
can be expected to be a general phenomenon observable also in BN
nanotubes and many other isotopically disordered confined systems.

\begin{acknowledgments}
We thank Niels Vandecasteel for his help with Fig.~\ref{fig:types}.
This work was supported by the ANR-PNANO2008 (project ACCATTONE).
\end{acknowledgments}

%
%


%
%
\end{document}